\documentclass[rmp,aps,amsfonts,amsmath,amssymb,nofootinbib,reprint,superscriptaddress]{revtex4-2}
\usepackage{graphicx}
\usepackage{bm}
\usepackage{hyperref}
\hypersetup{
    colorlinks,
    linkcolor={red!50!black},
    citecolor={blue!50!black},
    urlcolor={blue!80!black}
}

\usepackage{natbib}
\usepackage{float}
\usepackage{xcolor}

\graphicspath{{figures/}}
\begin{document}
\title{ Reconciling strange metal transport in \texorpdfstring{CeCoIn$_5$}{CeCoIn5} through the difference of optical and cyclotron effective masses}
\author{J. Wang}
 \affiliation{National High Magnetic Field Laboratory, Los Alamos National Laboratory, Los Alamos, NM 87545}
 \affiliation{William H. Miller III Department of Physics and Astronomy, Johns Hopkins University, Baltimore MD}

\author{Zhenisbek Tagay}
 \affiliation{William H. Miller III Department of Physics and Astronomy, Johns Hopkins University, Baltimore MD}

  \author{Liyu Shi}
 \affiliation{William H. Miller III Department of Physics and Astronomy, Johns Hopkins University, Baltimore MD}

\author{Jiahao Liang}
\affiliation{William H. Miller III Department of Physics and Astronomy, Johns Hopkins University, Baltimore MD}
 
 \author{Nghiep Khoan Duong}
 \affiliation{Department of Physics, Cornell University, Ithaca, New York 14853}

 \author{Yi Wu}
 \affiliation{Department of Physics, Cornell University, Ithaca, New York 14853}

 \author{P. M. T. Vianez}
  \affiliation{Institute for Materials Science, Los Alamos National Laboratory, Los Alamos, New Mexico, 87545}

 \author{F. Ronning}
 \affiliation{Institute for Materials Science, Los Alamos National Laboratory, Los Alamos, New Mexico, 87545}

  \author{D. G. Rickel}
\affiliation{National High Magnetic Field Laboratory, Los Alamos National Laboratory, Los Alamos, NM 87545}

 \author{Darrell G. Schlom}
 \affiliation{Department of Materials Science and Engineering,
Cornell University, Ithaca, New York 14853}
 \affiliation{Leibniz-Institut f{\"u}r Kristallz{\"u}chtung, 12489 Berlin}
  \affiliation{Kavli Institute at Cornell for Nanoscale Science, Ithaca, New York 14853}
 
 \author{K.M. Shen}
 \affiliation{Department of Physics, Cornell University, Ithaca, New York 14853}
 \affiliation{Kavli Institute at Cornell for Nanoscale Science, Ithaca, New York 14853}

 \author{S. A. Crooker}
\affiliation{National High Magnetic Field Laboratory, Los Alamos National Laboratory, Los Alamos, NM 87545}

\author{N.P. Armitage}%
 \email{npa@jhu.edu}
 \affiliation{William H. Miller III Department of Physics and Astronomy, Johns Hopkins University, Baltimore MD}
  \affiliation{Canadian Institute for Advanced Research, Toronto, Ontario M5G 1Z8}

\date{\today}

\maketitle

\textbf{The strange metal behavior in cuprate superconductors - characterized by linear in temperature resistivity and anomalous Hall transport - stands in stark contrast to the expectation of conventional Fermi liquid (FL) theory. Remarkably, the similar transport behavior has also been observed in the heavy fermion metal CeCoIn$_5$, whose d-wave superconducting ground state and strong antiferromagnetic fluctuations draw parallels to the cuprates. Here we have investigated the optical conductivity of the strange metal state of CeCoIn$_5$ over a wide magnetic field range using time-domain THz spectroscopy (TDTS).  Using unique high-field THz spectroscopy we have shown that the current relaxation rate scales approximately as T$^2$, giving evidence for a hidden Fermi liquid state over a large field range. This result can be reconciled with linear in T resistivity with the realization that heavy quasiparticles have an optical mass that scales roughly like 1/T.  This optical mass contrasts with the mass that characterizes cyclotron motion, which does not suffer the same large temperature dependent renormalization. Although by itself anomalous, this allows one to understand a number of other phenomena in CeCoIn$_5$ that have been taken to be signatures of strange metals, including the coexistence of a conventional T$^2$ dependence of the cotangent of the Hall angle with the linear in T resistivity, which with our observation also reflects FL-like physics.}

The heaviest electronic effective masses $m^*$ in condensed matter physics can be found in heavy-fermion materials, which is evidenced by a large low temperature Sommerfeld coefficient in the heat capacity~\cite{andres19754}. In the optical conductivity spectrum this coherent Kondo state manifests itself as the opening of the hybridization gap, a reduction in the spectral weight at frequencies below the gap, and frequently the appearance of a narrow Drude-like peak at low frequencies~\cite{RevModPhys.71.687,Basov2011471,PhysRevB.85.155105}. The latter two indicate the enhancement of the optical mass $m^*$ upon application of the $f$-sum rule.  However, even within Fermi liquid (FL) theory different experimental probes can be sensitive to masses in different ways.  Many experimental probes -- such as heat capacity, optical spectroscopy, quantum oscillation, and angle-resolved photoemission~\cite{pines1967theory} - are sensitive to both electron-electron and electron-phonon interactions, whereas (in a Galilean invariant system) compressibility and spin susceptibility~\cite{pines1967theory} are sensitive only to electron-electron interaction. Charge transport provides yet another perspective, since the relevant mass renormalizations are those arise from primarily momentum non-conserving scattering.


\begin{figure*}[t]
\centering
\includegraphics[width=0.7\textwidth]{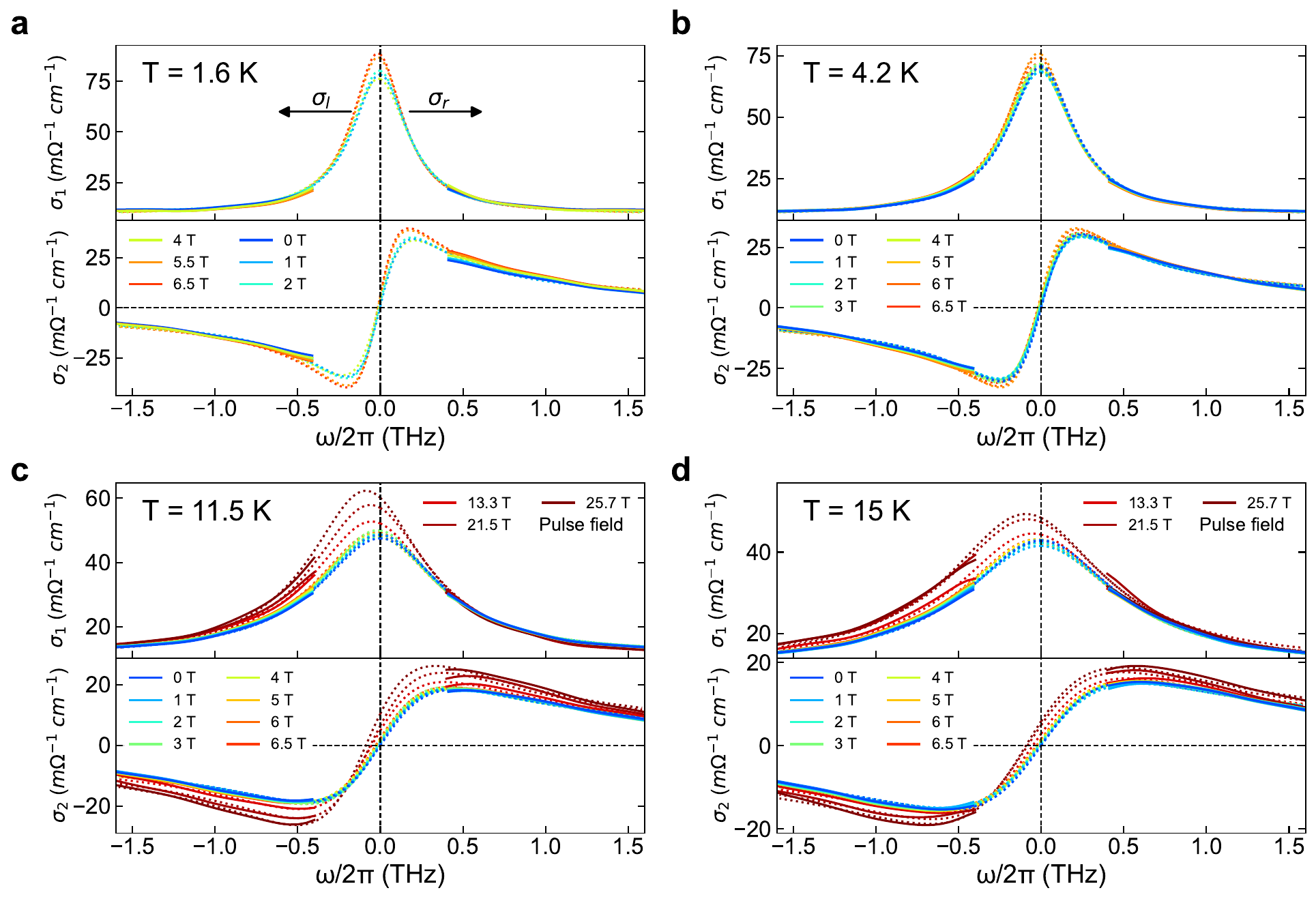}
\caption{\textbf{THz conductivity of the CeCoIn$_5$ thin film.} Real and imaginary part of the THz conductivity ($\sigma_1$ and $\sigma_2$, respectively) in the circular basis at the temperatures and in the magnetic fields indicated, showing systematic electron-like cyclotron resonance. a) b) at low temperature in the JHU setup. c)  d) higher temperature data including the pulsed field data taken at LANL. The dashed lines are fits using two Drude terms. Superconducting phase data (1.6K, below 2T) are included for completeness.}\label{fig1}
\end{figure*}

Cyclotron resonance (CR) -- the periodic circulating motion of charge carriers -- is however more subtle and requires additional consideration. As Kohn elegantly demonstrated~\cite{kohn1961cyclotron}, the cyclotron mass of a Galilean invariant system is unaffected by electron-electron interaction, because CR is a collective excitation determined solely by the center of mass motion of the electrons. In real systems, various mechanisms can violate Kohn's theorem: significant Umklapp scattering in systems with high carrier densities breaks the electron Galilean invariance; electron-impurity scattering and Kondo interactions modify the cyclotron response by coupling electrons' center of mass coordinate to their internal coordinate; nonparabolic band dispersion allows cyclotron mass renormalization by electron-electron interaction~\cite{allen1982cyclotron,kanki1997theory,macdonald1989cyclotron,hu1988memory}. These mechanisms can modify the cyclotron mass in ways different from the usual FL renormalizations, making CR a complementary and unique probe of mass and electron-electron interactions~\cite{kanki1997theory}.  The cyclotron mass $m_c$ is defined by the inverse proportionality to the cyclotron resonance frequency in a magnetic field, by the simple relation $\omega_c=eB/m_c$.  It is challenging to resolve the resonance of heavy carriers when strong scattering suppresses sharp spectral features.  High precision, high magnetic field measurements are therefore required for strongly interacting systems, and only a few studies exist due to the experimental challenges~\cite{post2021observation,legros2022evolution,wang2025thz}.

\begin{figure*}[t]
\centering
\includegraphics[width=0.75\textwidth]{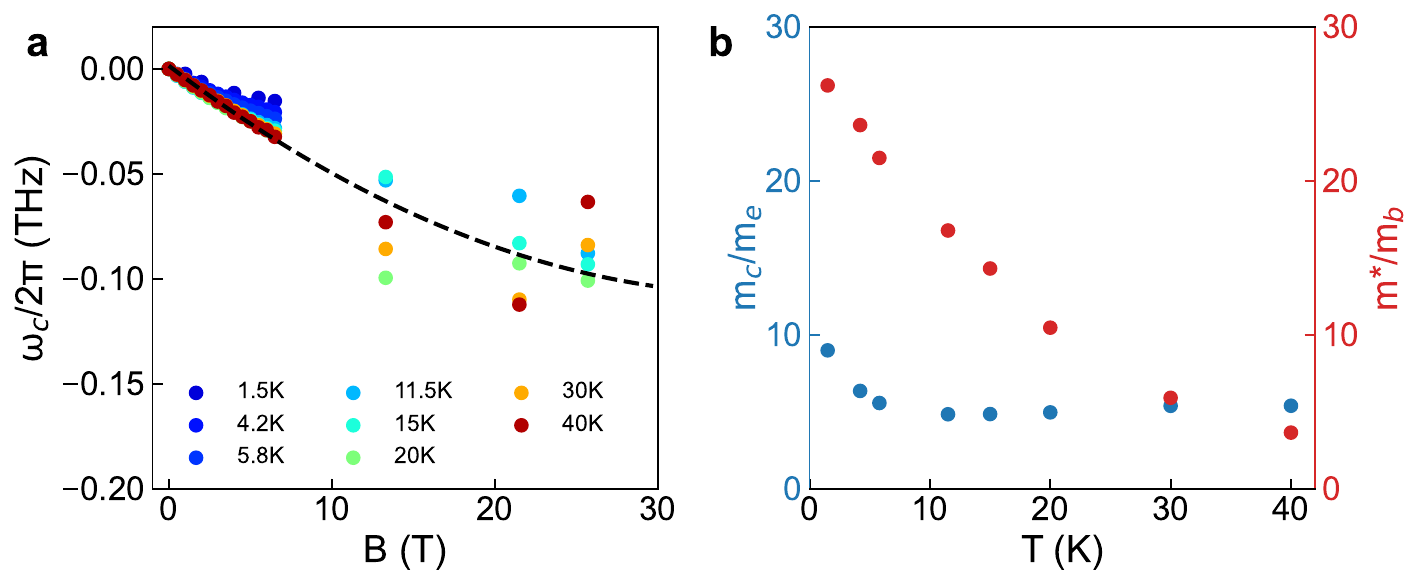}
\caption{\textbf{Cyclotron frequency $\omega_c$ and cyclotron mass $m_c$.} a) Field dependence of $\omega_c$ at different temperatures. $m_c$ is field independent in the low field region as $\omega_c$ is linear in $B$. In high fields $m_c$ increases as indicated by the decreased slope. The dashed line is a guide to the eye. b) Temperature dependence of the optical effective mass $m^*$ extracted from low field EDM analysis and $m_c$ from a linear fit to $\omega_c$ below 2T.}\label{fig2}
\end{figure*}


In this work, we perform high magnetic field THz polarimetry on the heavy fermion superconductor CeCoIn$_5$~\cite{Petrovic2001L337}, a widely studied system of particular interest because it has the highest $T_c$ among any non-Pu based heavy-fermion superconductor. Its normal state has a rich magnetic-field-tuned phase diagram: it is proximate to a magnetic field tuned quantum critical point (QCP) at 5 T and has been reported to have a non-FL normal state that persists up to 25K at low fields, and a conventional FL normal state that only emerges at low temperatures under strong magnetic fields above the QCP~\cite{paglione2003field}. The non-FL normal state is typified by a linear in T dependent resistivity, which is in conflict with the cotangent of the Hall angle that follows a more conventional T$^2$ dependence, and a strong violation of Kohler's rule~\cite{nakajima2004normal}. This is reminiscent of the non-FL transport in the cuprates~\cite{ando2004evolution}. However, we show that the current relaxation rate in fact scales like T$^2$, giving evidence for a hidden Fermi liquid state over a large field range. In this picture, the FL behavior in longitudinal transport is obscured by a strong quasiparticle optical mass renormalization that approximately follows $1/T$.  In addition, we find that this optical mass contrasts with the mass that characterizes cyclotron motion, which does not suffer the same large temperature dependent renormalization. The distinct temperature evolution of the optical mass and cyclotron mass naturally reconciles the "non-FL" behavior in the magnetotransport of CeCoIn$_5$, highlighting again the significant role effective masses play in unconventional charge transport behaviors.

TDTS polarimetry was performed on 60 nm CeCoIn$_5$ film grown on a MgF$_2$ substrate.  Complex optical conductivity was measured up to 26 T in a novel pulsed magnetic field setup at LANL~\cite{post2021observation,legros2022evolution}. High-field data was supplemented with high-precision THz polarimetry up to 6.5T at JHU~\cite{tagay2024high}. CR is most clearly revealed by plotting the optical conductivity in the circular polarization basis ($\sigma_r$, $\sigma_l$) as they are the eigenbasis of cyclotron motion.   We use a convention where the frequency runs positive and negative to represent right- and left-hand polarized light.  Details of the circular basis analysis are discussed in {\bf Methods}.  As shown in Figure \ref{fig1} (and discussed previously~\cite{PhysRevB.72.045119,PhysRevB.65.161101,shi2025low}), at least 2 Drude oscillators are needed to fit the real and imaginary conductivities simultaneously at the lowest temperatures.  A sharp Drude peak with small scattering rate and a very broad Drude oscillator that gives almost only a real contribution to the conductivity together give a rough emulation of the conductivity. This is a typical optical conductivity for a heavy fermion system at low temperatures~\cite{PhysRevB.65.161101,PhysRevB.85.155105,PhysRevB.72.045119,PhysRevB.93.085104,RevModPhys.71.687}.  We show in Fig. \ref{fig1} an electron-like cyclotron resonance as indicated by a subtle, but distinct shifting of the conductivity to negative frequency at a few different temperatures. The conductivity can still be fitted by the two Drude model, with a cyclotron shift included in the sharp Drude component, while the broad Drude component is not very sensitive to magnetic field.   The cyclotron frequency is the central resonance of the sharp Drude component.  In the circular polarized basis the Drude model has the simple spectral dependence where the Drude peak simply shifts by the cyclotron frequency $\omega_c = \frac{eB}{m_c}$ e.g.

\begin{align}
\sigma_{r,l}(\omega) = \frac{Ne^2}{m^*} \frac{\tau^*}{1-i(\omega - \omega_c)\tau^*}.
\label{Drude}
\end{align}

In conventional Drude theory the mass $m^*$ in the prefactor of this expression is the same as the mass in the cyclotron resonance frequency $m_c$.  Here we allow -- and our data demands -- them to be different.  The simultaneous fitting for the real and imaginary parts of both left- and right-circular conductivity puts strong constraints on the fit, DC Hall resistivity is used to further constraint the fit to fully capture the subtle field and temperature dependence in the magnetotransport. These constraints enable reliable analysis of the data despite the fact that we can not take data near $\omega = 0$.  For a complicated multi-band system like CeCoIn$_5$, the 2 Drude fit should not be necessarily interpreted as the optical conductivity of two types of charge carriers, but rather as a minimal parametrization of the complex frequency dependent data.  Physically, the small low frequency peak characterizes the fact that a large portion of the current is carried by a component that has a significant inertial response, which in the language of individual charge carriers can be called a large mass.

The field dependence of $\omega_c$ from this parametrization is shown in Figure \ref{fig2}a). We observed $\omega_c$ linear in $B$ up to approximately 2T which gives an almost field independent cyclotron mass at low fields at all displayed temperatures, as shown in Figure \ref{fig2}b. At higher fields, $\omega_c$ deviates from the linear in $B$ behavior and starts to saturate. If the relation $\omega_c = \frac{eB}{m_c}$ remains valid, this implies an enhanced $m_c$ at high fields. This is in line with the field suppressed Hall coefficient in CeCoIn$_5$\cite{hundley2004anomalous,nakajima2004normal,maksimovic2022evidence}.

\begin{figure*}[t]
\centering
\includegraphics[width=0.65\textwidth]{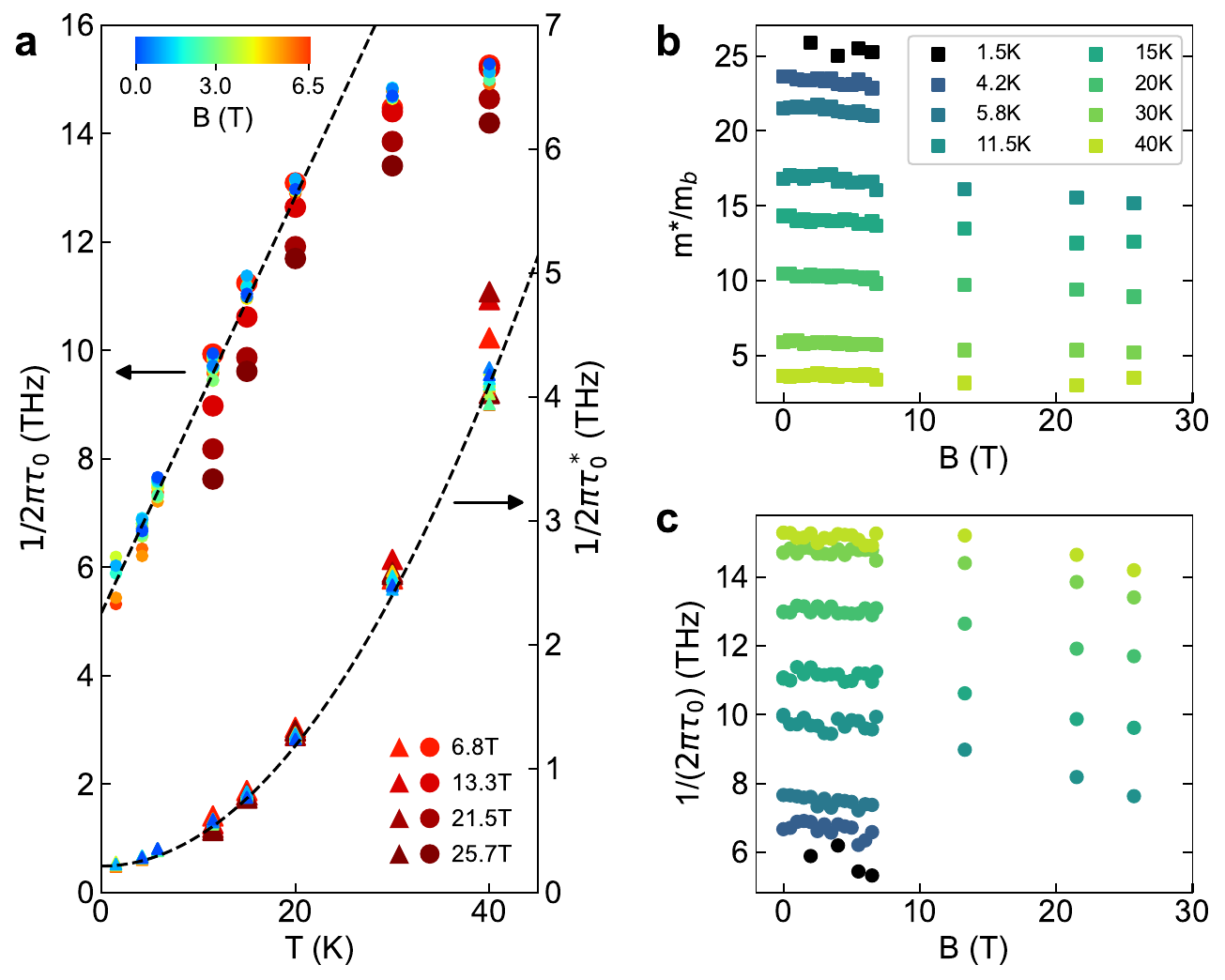}
\caption{\textbf{Zero-frequency scattering rates and effective masses extracted from EDM analysis.} a) Temperature dependence of zero-frequency value for scattering rate $1/\tau_0$ and normalized scattering rate $1/\tau^*_0$. $1/\tau_0$ is linear in $T$ while $1/\tau^*_0$ follows $T^2$, as indicated by the dashed lines. Larger symbols denote pulse field data. b), c) Field dependence of the effective mass and the scattering rate $1/\tau_0$ at zero frequency.}\label{fig3}
\end{figure*}

As has been widely used in optical spectroscopy in zero magnetic field, important insights about electron interactions can be gained by analyzing a non-Drude conductivity spectrum using Extended Drude Model (EDM) or memory function approaches, where a frequency dependent mass and scattering rate can be extracted by inverting the complex conductivity~\cite{Basov2011471,allen1982cyclotron,hu1988memory}. In a magnetic field, similar ability to extract the frequency dependent scattering rate and mass has been challenging because of the inability of conventional Fourier transform infrared spectroscopy to measure the complex conductivity in field directly as it requires the ability to measure the complex transmission function in circularly polarized basis.  Older experiments and analyses and attempts at EDM in field could only approximate the complex $\sigma_{xx}(\omega)$ and were not sensitive to the complex $\sigma_{xy}(\omega)$~\cite{allen1982cyclotron}.  This opportunity is uniquely allowed by TDTS polarimetry. Following Quinn and others one can write an expression that follows Eq.~\ref{Drude} in that quantities $m^*$, $\tau^*$ and $\omega_c$ become frequency dependent~\cite{ting1976infrared,ting1977theory,hu1988memory}.   See Supplementary Information for further discussion.  Note again that an effective mass appears in this expression in two places: in the pre-factor $\frac{Ne^2}{m^*}$ and in $\omega_c$.   As discussed above, in the conventional Drude treatment these masses are the same. As discussed below, we must allow them to be independent quantities here.  We can invert the optical conductivity to get the frequency dependent mass and scattering rate:

\begin{align}
\frac{1}{\tau(\omega)}&= \epsilon_0 \omega_p^2 \text{Re}[\frac{1}{\sigma_{r,l}(\omega)}] , \\
\frac{m^*(\omega)}{m_b} &= - \epsilon_0 \frac{\omega_p^2}{(\omega - \omega_c)}  \text{Im}[\frac{1}{\sigma_{r,l}(\omega)}].
\end{align}

Here $m^*(\omega \rightarrow 0 )$ is the effective mass, $\frac{1}{\tau^*(\omega)}=\frac{1}{\tau(\omega)}\frac{m_b}{m^*(\omega)}$ is the scattering rate $\frac{1}{\tau}$ normalized by the mass enhancement factor $\frac{m^*}{m_b}$, with $m_b$ the band mass.  If the frequency dependencies are weak enough $\frac{1}{\tau^*(\omega \rightarrow 0)}$ would be the approximate width of the low frequency Drude peak and $\frac{\pi}{2} \epsilon_0 \omega_p^2 = Ne^2/ m_b$ its total spectral weight integrated over the low frequency band-width. In what follows we take $\omega_p$ to be 29,700 cm$^{-1}$ (892 THz) from  the integral of the spectral weight up to 2500 cm$^{-1}$ ~\cite{lee2023temperature}.  In Figure \ref{fig2}b) we compare the temperature dependence of the cyclotron mass $m_c$ from the Drude fits above to the low frequency limit of inertial mass $m^*$.  Mass renormalization modifies $m_c$ and $m^*$ in distinct ways. While $m^*$ is strongly enhanced at low temperatures due to the onset of the strongly interacting coherent Kondo state, $m_c$ has a much weaker temperature dependence.   Fig.~\ref{fig2}b.) is the key result of this work.

\begin{figure*}[t]
\centering
\includegraphics[width=0.65\textwidth]{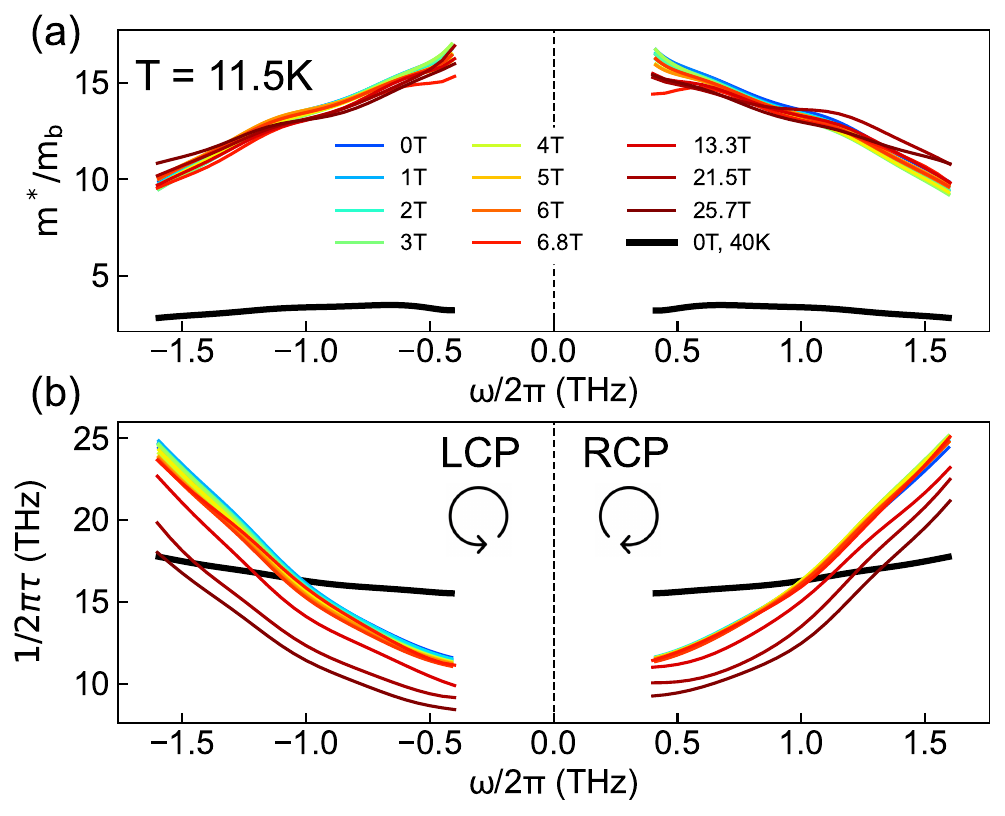}
\caption{\textbf{Frequency dependent effective masses and scattering rates extracted from EDM analysis} a) Effective masses, frequency dependent mass is an evidence for strong electronic correlation. No polarization or field dependence are observed. b) Scattering rate, favored (left circularly polarized) polarization has a lower scattering rate. The overall magnitude of the scattering rate decreases in field, consistent with the field suppressed DC resistivity.}\label{fig4}
\end{figure*}

In Fig. \ref{fig3}a we compare the zero-frequency scattering rates $\frac{1}{\tau^*_0}=\frac{1}{\tau^*( \omega \rightarrow 0 )}$ and $\frac{1}{\tau_0}=\frac{1}{\tau( \omega \rightarrow 0 )}$.   Whereas the former shows a notably T$^2$ like dependence, the latter shows a linear-in-T dependence.   The former is the rate of scattering of the heavy charge carriers where the resistivity $\rho_{DC} \propto m^*/\tau^*_0$, whereas the latter is a formal quantity whose $\rho_{DC} \propto m_b/\tau_0$.  Figure \ref{fig3} shows the value for both scattering rates as a function of temperatures and magnetic fields. Little field dependence is observed at the low fields; at high fields, both $m^*$ and $1/\tau_0$ decrease roughly proportionally, leaving their ratio $1/\tau^*_0$ unchanged with field.   CeCoIn$_5$ in this temperature range is commonly claimed to be a non-FL due to its strange-metal-like linear in T resistivity. However, $1/\tau_0^*$ can be well described by a FL-like scattering $\frac{1}{\tau^*(\omega \rightarrow 0)}=B+AT^2$, as indicated by the dashed line.  Although it is roughly the width of the narrow Drude peak in the spectra, we emphasize that $1/\tau^*(\omega \rightarrow 0)$ can only be reliably obtained by extrapolating the EDM analysis at finite frequency to DC. The observed $T$ linear dependence in $\rho_{DC}$ arises from the strong temperature dependence of the effective mass, $m^*/m_b$ approximately follows a $1/T$ dependence at low temperatures that scales the underlying T$^2$ quasiparticle scattering rate into a T linear resistivity, hiding the FL nature of the quasiparticles.   This phenomena -- previously found at zero magnetic field~\cite{shi2025low} -- is established now for the whole measured magnetic field range.

The existence of two different masses immediately gives an interesting expression for the DC Hall coefficient from the zero frequency limit of Eq.~\ref{Drude}, $R_H = \frac{1}{Ne}\cdot\frac{m^*}{m_c}$. This provides a natural explanation for the temperature dependent Hall coefficient observed in CeCoIn$_5$. In contrast to the conventional single band case where the Hall coefficient depends only on $N$, it is now dressed by the ratio of the two masses i.e. they do not cancel.  As $m_c$ is relatively temperature independent, the Hall coefficient is determined primarily by the effective mass $m^*$, except for below 5K the cyclotron mass $m_c$ starts to increase and causes a decrease in $R_H$.

Our observations also naturally account for the conflict between the linear-in-T resistivity and the more conventional T$^2$ dependence of the cotangent of the Hall angle \cite{hundley2004anomalous,nakajima2004normal}. Eq.\ref{Drude} predicts cot $\theta_H = \frac{\sigma_{xx}}{ \sigma_{xy}  }= \frac{1}{ \omega_c \tau^*_0  }$.  Because the cyclotron mass $m_c$ is relatively temperature independent in this range, the temperature dependence of our cotangent of the Hall angle acquires the same temperature dependence as $1/\tau^*_0$, which is the more conventional FL-like T$^2$.  At the same time, we emphasize that the appearance of two different masses remains anomalous. 

Next, we analyze the frequency dependent mass and scattering rates measured at the lowest temperature (11.5K) using the pulse field setup. They are shown in Figure \ref{fig4}, plotted in the same circular basis as the conductivity that represents the left- and right-handed motion of the electrons. Compared to the 40K data (black), both $\frac{m^*(\omega)}{m_b}$ and $\frac{1}{\tau}(\omega)$ at 11.5K acquires a strong frequency dependence, reflecting the strong electronic correlations in the coherent Kondo state. The renormalization of these Kramers-Kronig related quantities are largest at low frequencies.  The enhancement is much smaller and both quantities should approach their frequency-independent unnormalized values at high-enough temperatures and frequencies \cite{dressel2002electrodynamics}. This behavior is most clearly seen in $\frac{m^*(\omega)}{m_b}$, which decreases with frequency and approaches unity at sufficiently high temperature.  At low T, $\frac{1}{\tau (\omega)}$ follows a power-law frequency dependence, with an exponent close to the $\omega^2$ behavior expected for a Fermi-liquid.  $\frac{m^*(\omega)}{m_b}$ does not have a significant field dependence up to the highest field. $\frac{1}{\tau(\omega)}$ shows two features in magnetic field. First, its overall magnitude decreases with field, consistent with the decreasing DC resistivity in magnetic field~\cite{paglione2003field}. Second, when extrapolated to lower frequency, the minimum of $\frac{1}{\tau(\omega)}$ (as well as $\frac{1}{\tau^*(\omega)}$) no longer occurs $\omega=0$ but instead shifts to $\omega=\omega_c$. Phenomenologically this is a consequence of the fact that $\frac{1}{\tau(\omega)}$ is the inversion of a Drude-like oscillator centered around $\omega_c$. This differs from the usual FL theory, in which $\omega^2$ scattering rates arises from the phase space considerations that a quasiparticle excited above $\epsilon_F$ by $\hbar \omega$ can only scatter with quasiparticles $\hbar \omega$ below $\epsilon_F$ and can only scatter into states $\hbar \omega$ above $\epsilon_F$.

 
Our observation of different optical and cyclotron masses allows our data to resolve a long outstanding mystery in CeCoIn$_5$: how to reconcile the T-dependent resistivity and the T$^2$-dependent cotangent of the Hall angle. They both reflect the T$^2$ dependence of the scattering rate of heavy-quasiparticles, however the resistivity has a factor of $m^*$ that roughly follows $1/T$, whereas the cotangent of $\theta_H$ has a factor of  $m_c$, which is relatively temperature independent.  A simple magnetotransport extension of the Drude model (See Supplemental Information) allows one to understand these two different masses as either two different masses or two different scattering rates for longitudinal and Hall transport as discussed long ago by Anderson for the cuprates~\cite{anderson1991hall}.  Although we present this simple phenomenological scenario to understand our data, it would be interesting to understand if models for fractionalized phases~\cite{ioffe1989gapless,senthil2004weak} that have been used successfully to model the DC Hall effect could be used to describe this low frequency Hall data~\cite{maksimovic2022evidence}.  It may be possible that this allows the understanding of the violation of Kohler's rule~\cite{nakajima2004normal} as the simple magnetotransport model does not.  It would be interesting to investigate this physics in other heavy-fermion systems that exhibit linear in T resistivity and cot$\theta_H$$\propto$T$^2$  suhc as YbRh$_2$Si$_2$ and CeCu$_{5.9}$Au$_{0.1}$~\cite{schroder2000onset,gegenwart2003tuning,custers2003break,paschen2004hall}.  We also note that although the present linear in T dependence of the resistivity and T$^2$ dependence of cot $\theta_H$ is phenomenologically similar to the cuprates, the resolution here is unlikely to apply to that case, as cuprates do not appear to have a strongly temperature dependent $m^*$ (despite different masses from cyclotron resonances and heat capacity measurements ~\cite{legros2022evolution}). This may point to a disappointing but nevertheless true lack of universality in the explanation for strange metals.

\medskip

\small{\textsf{\textbf{ACKNOWLEDGMENTS}}}

\medskip

We would like to thank Piers Coleman for important conversations.  This project at JHU was supported by the Gordon and Betty Moore Foundation EPiQS Grant No. GBMF-9454, and by the United States Department of Defense under grant No. W911NF2120213.  NPA had additional support from the Quantum Materials program at the Canadian Institute for Advanced Research.  At Cornell this work was supported by the Gordon and Betty Moore Foundation's EPiQS initiative through Grant Nos. GBMF3850 and GBMF9073.  Additional support for materials synthesis was provided by the National Science Foundation through Grant No. DMR-2104427 and Cooperative Agreement No. DMR-2039380 through the Platform for the Accelerated Realization, Analysis, and Discovery of Interface Materials (PARADIM).  Work at LANL was supported by the US Department of Energy (DOE) through the Basic Energy Sciences "Science of 100T" program.  The National High Magnetic Field Lab is supported by the NSF under grant DMR-2128556, the State of Florida, and the US DOE.

\medskip

\small{\textsf{\textbf{METHODS}}}

\medskip

CeCoIn$_5$ thin films were grown on 10 mm $\times$ 10 mm  MgF$_2$ substrates by molecular-beam epitaxy (MBE) at Cornell. Growth was performed in a Veeco GEN10 system with elemental sources of cerium (99.99$\%$), cobalt (99.99$\%$) and indium (99.99$\%$) in a background pressure better than $5 \times 10^{-9}$ torr, following a similar procedure detailed by Mizukami et al.~\cite{Mizukami_2011}. The elemental fluxes were calibrated using a quartz crystal microbalance and x-ray reflectivity measurements of film thicknesses before growth. The MgF$_2$ substrates were first pre-annealed at 750$^\circ$  before growth until a clear reflection high-energy electron diffraction (RHEED) pattern was observed, and then lowered to 375$^\circ$  for the film deposition. The growth rate was 1.5-2.5 nm/min and monitored in real-time using RHEED. Samples were characterized post-growth using x-ray diffraction and electrical resistivity measurements. 

The thickness of the films were limited at 60 nm to have sufficient THz transmission. Residual resistivity ratios of approximately 5 are typical, which is lower than single crystals. The Hall effect is more reminiscent of CeCoIn$_5$ single crystals under pressure, which shows roughly the same phenomenology~\cite{nakajima2006evolution}. See more details in Supplementary Information.

The sample was kept in a vacuum or helium environment after growth and before measurement to minimize oxidization. A bare MgF$_2$ substrate was used as a reference in the THz measurement.  The THz spectrum from 0.2 - 1.8 THz was measured on a fiber-coupled Toptica time-domain THz spectrometer modified to allow both $x$ and $y$ components of the THz electric field to be measured simultaneously. By dividing the Fourier transform of electric field transmission through the sample by the Fourier transform of transmission through the bare MgF$_2$ reference substrate, the complex transmission $T(\omega)$ is obtained and the complex optical conductivity ($\sigma = \sigma _1 + i \sigma _2$) can be calculated from $T(\omega)$ by the usual expression $T(\omega) = \frac{1+n}{1+n+\sigma d Z_0}e^{\frac{i\omega \Delta L(n-1)}{c}}$.  Here $n$ is the substrate index of refraction, $d$ is the film thickness, $\Delta L$ is a correction factor that accounts for thickness differences between the reference and sample substrates, and Z$_0$ is the impedance of free space (377 $\Omega$). Optical conductivities are measured in the linear polarization basis, the circular basis values are obtained by transformation from linear basis using $\sigma_{r,l}=\sigma_{xx}\pm i\sigma_{xy}$.


\bigskip

\bibliography{CeCoIn5_database.bib}

\newpage

 \textbf{SUPPLEMENTARY INFORMATION}

\bigskip

\textbf{I. Discussion on extended Drude models in a magnetic field}

\bigskip

Following the definition in Refs.~\cite{ting1976infrared} and incorporating a frequency dependent memory function $M(\omega) = M_1(\omega) + i M_2(\omega)$, one can write

\begin{align}
\sigma_{r,l}(\omega)
&= \frac{i N e^2}{m_b}\,\frac{1}{\omega - \omega_c + M(\omega)} \\
&= \frac{N e^2}{m_b}\,\frac{1}{M_2 - i(\omega - \omega_c + M_1)} \\
&= \frac{N e^2}{m_b\left(1 + \frac{M_1}{\omega}\right)}
   \,\frac{1}{\dfrac{M_2}{1 + \frac{M_1}{\omega}}
   - i\left(\omega - \dfrac{\omega_c}{1 + \frac{M_1}{\omega}}\right)} 
\end{align}

Where the frequency argument runs over positive and negative frequencies to express the response to right and left-hand polarized light.  We can identify in the usual way a correspondence of the memory functions with the effective mass and scattering rate e.g. $m^*/m_b  = 1+\frac{M_1}{\omega}$ and $\frac{1}{\tau} = M_2$.   Furthermore we can define $\frac{1}{\tau^*} = \frac{1/\tau }{ m^*/m_b} $, $\omega_c^* = eB/m^*$, and $\frac{\omega_p^{*2}}{4 \pi } = \frac{Ne^2}{m^*}$, where the assumption is that all these quantities (except $e$ and $N$) are frequency dependent. We then have

\begin{align}
   &= \frac{N e^2}{m^*}\,
   \frac{1}{\dfrac{1}{\tau^*} - i(\omega - \omega_c^*)}, \\
&= \frac{\omega_p^{*2}}{4\pi}\,
   \frac{\tau^*}{1 - i(\omega - \omega_c^*)\tau^*}.
\end{align}

This a standard and natural extension of the usual Drude magneto-conductivity expression in Eq. 1 that assumes that the same renormalized mass appears in the spectral weight prefactor and in the denominator to cyclotron frequency.   As discussed in the main text, such an expression does not fit our data.   We are forced to adopt the expression 

\begin{align}
\sigma_r(\omega) = \frac{N e^2}{m^*}\,
   \frac{1}{\dfrac{1}{\tau^*} - i(\omega - \omega_c)},
   \label{memorymod}
\end{align}

where $\omega_c = eB/m_c$ e.g. the two masses $m^*$ and $m_c$ are different.  As discussed in the main text, a reminiscent phenomenology to this has been observed in cuprate~\cite{carrington1992temperature,kaplan1996normal,zheleznyak1998phenomenological}.

It is challenging to motivate this expression, but one way it can be done is to assume Hall currents decay at a different rate than conventional currents.  Following Ref.~\cite{romero1992cyclotron}, we can write an equation in terms of velocities for a charge in the presence of external electric and magnetic fields

\begin{equation}
\tau \frac{d v_i}{d t} + v_i = \sum_{j=1}^{3} \mu_{ij}\,\xi_{ij}.
\label{EqMotion}
\end{equation}

For a magnetic field in the $z$ direction the ``electric field" matrix is

\begin{equation}
\xi_{ij} =
\begin{pmatrix}
E_x & \dfrac{v_y B}{c} \\
-\dfrac{v_x B}{c} & E_y
\end{pmatrix}.
\label{pmatrix}
\end{equation}

And the mobility tensor is

\begin{equation}
\mu_{ij} = e
\begin{pmatrix}
\tau_{tr}/m^* & \tau_{H}/m_{c} \\
\tau_{H}/m_{c} & \tau_{tr}/m^*
\end{pmatrix}.
\label{mumatrix}
\end{equation}

Note the order of the indices in the multiplication in the sum of Eq. \ref{EqMotion} and is not matrix multiplication, but instead a product of matrix elements.  Eq.~\ref{EqMotion} can be solved with the usual substitutions and in response to circularly polarized light $E = E_x \pm iE_y$ to get conductivity

\begin{equation}
\sigma_{\pm}
=  \frac{N e^2}{m_{tr}}
\frac{1 }
{ \frac{1}{\tau_{tr} } - i( \omega - \omega_c  \frac{\tau_H}{\tau_{tr} } ) }.
\label{Romero}
\end{equation}

With the assumption that $\tau_H = \tau_{tr}$ Eqs. \ref{memorymod} and \ref{Romero} straightforwardly correspond to each other with the two different masses.  Alternatively, one could imagine the two masses in Eq.~\ref{mumatrix} are the same, but the effective cyclotron frequency is renormalized by the ratios in scattering rates.  With appropriate substitutions of $\frac{1}{\tau_{tr}} \propto$ T and $\frac{1}{\tau_{H}}\propto$ T$^2$ this expression also reproduces the cot$\theta_H$$\propto$T$^2$ dependence.  This expression does not explain the violation or anomalous scaling of Kohler's rule~\cite{nakajima2004normal}.

However, some caveats must be made.  The starting equation of motion is unconventional as typically one would write an equation of motion like Eq. \ref{EqMotion} with a force term by itself and not multiplied by a mobility.  It normally would not matter if the mobility matrix was diagonal, but in this case is necessary to write it as such because in this formulation one needs to distinguish the part of the current that comes from a magnetic field and the part that comes from an electric field and postulate that they decay at different rates.  In this formulation, the effective force term violates Lorentz invariance as the E and B terms have different prefactors.  

Another way to motivate a similar phenomenology is to give different regions of the momentum space different scattering rates, as has been done in the cuprates~\cite{zheleznyak1998phenomenological,nakajima2006evolution}.  If the scattering rate is assumed to have a linear temperature dependence on flat parts of the Fermi surface and a quadratic scattering rate in the corners, then the data can be accounted for. The longitudinal conductivity is dominated by the contribution from the flat regions, whereas the sharply curved corners control the Hall conductivity.

\bigskip

\textbf{II. Hall data on the CeCoIn$_5$ film}

\bigskip

\setcounter{figure}{0}
\renewcommand{\thefigure}{S\arabic{figure}}

\begin{figure*}[t]
\centering
\includegraphics[width=0.65\textwidth]{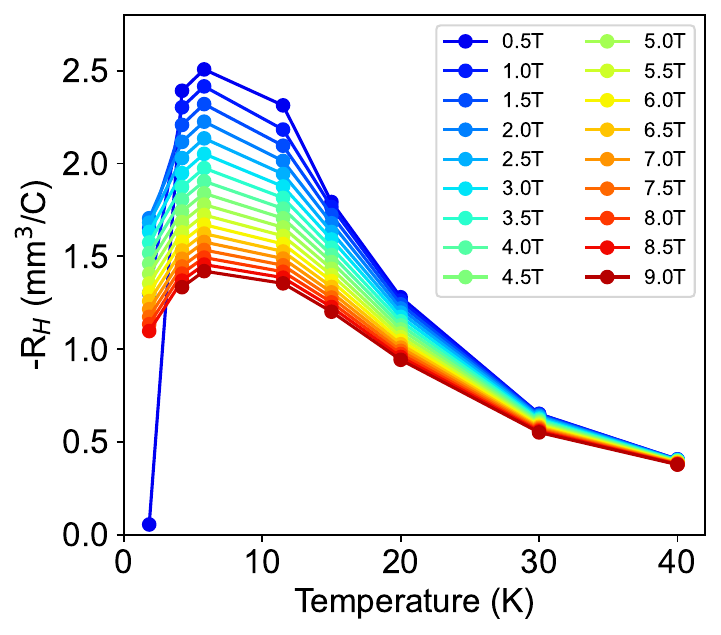}
\caption{\textbf{DC Hall coefficient} obtained by electrical transport measurement on a similarly prepared film.}\label{figs1}
\end{figure*}

The DC Hall coefficient was measured on a different as-grown CeCoIn$_5$ thin film, because the film used for THz measurements was cleaved and prepared for pulse field experiments. Fig.~\ref{figs1} shows the DC Hall coefficient measured under various magnetic fields, defined as $R_H(B)=\frac{\rho_{xy}}{B}$. Although the magnitude of $R_H$ is smaller than that reported for single crystals - presumably caused by strain that suppresses the $R_H$ as studied in single crystals\cite{nakajima2006evolution} - the two key features, namely the low temperature enhancement and the field suppression of $R_H$, are both clearly observed. This indicates that the thin film preserves the essential heavy fermion physics of its bulk form. The 2-Drude fit is under the constraint that its DC value gives a Hall coefficient that matches the transport data, to ensure that the parametrization of the THz data can be reasonably extrapolated to the DC limit.

\end{document}